\documentclass[aps,prl,twocolumn,showpacs,preprintnumbers,amsmath,amssymb,superscriptaddress]{revtex4}

\usepackage{graphicx}
\usepackage{dcolumn}
\usepackage{bm}

\def\ra{\rangle}
\def\la{\langle}
\def\dag{^\dagger}
\def\om{\hat \Omega}
\def\Tr{{\rm Tr}}
\def\cos{{\rm cos}}
\def\sin{{\rm sin}}
\begin{document}
\title{Exact Analysis of Entanglement in Gapped Quantum Spin Chains}

\author{Hosho Katsura}
\email{katsura@appi.t.u-tokyo.ac.jp}
\affiliation{Department of Applied Physics, the University of Tokyo,
7-3-1, Hongo, Bunkyo-ku, Tokyo 113-8656, Japan}

\author{Takaaki Hirano}
\email{hirano@pothos.t.u-tokyo.ac.jp}
\affiliation{Department of Applied Physics, the University of Tokyo,
7-3-1, Hongo, Bunkyo-ku, Tokyo 113-8656, Japan}

\author{Yasuhiro Hatsugai}
\email{hatsugai@pothos.t.u-tokyo.ac.jp}
\affiliation{Department of Applied Physics, the University of Tokyo,
7-3-1, Hongo, Bunkyo-ku, Tokyo 113-8656, Japan}

\date{\today}

\begin{abstract}
We investigate the entanglement properties of the valence-bond-solid states with generic integer-spin $S$. Using the Schwinger boson representation of the valence-bond-solid states
, the entanglement entropy, the von Neumann entropy of a subsystem, is obtained exactly 
and its relationship with the usual correlation function is clarified. The saturation value of the entanglement entropy, $2 \log_2 (S+1)$, is derived explicitly and is interpreted in terms of the edge-state picture. The validity of our analytical results and the edge-state picture is numerically confirmed. We also propose a novel application of the edge state as a qubit for quantum computation. 
\end{abstract}

\maketitle
Entanglement properties of quantum spin systems have been attracting much attention in quantum information theory and condensed matter physics.      
The entanglement entropy (EE), the von Neumann entropy of the reduced density matrix of a subsystem, is a measure to quantify how much entangled a many-body ground state is. 
Recently the EE has been used to investigate the nature of quantum ground states such as the quantum phase transition and topological/quantum order \cite{Vidal, Levin, Kitaev, Ryu, YH1}. 
Vidal {\it et  al.}\cite{Vidal} conjectured that the EE of a large block of spins in gapped spin chains reaches saturation while that in critical spin chains shows a logarithmic divergence.

In this Letter we study the EE of gapped quantum spin chains with arbitrary integer-spin.  
After the Haldane conjecture that integer-spin antiferromagnetic Heisenberg chains have a finite gap\cite{o3, Haldane}, Affleck, Kennedy, Lieb and Tasaki (AKLT) proposed the valence bond solid (VBS) state which enables us to understand ground state properties of the Haldane gap systems \cite{AKLT1,AKLT2}. 
The VBS is now attracting renewed interest from the viewpoint of quantum information theory. For example, universal quantum computation based on the VBS states has been proposed \cite{Verstraete and Cirac}. 

While the entanglement properties in $S=1$ VBS has been extensively studied in \cite{Cirac, Korepin}, we investigate the EE in generic VBS states with arbitrary integer spin $S$.
We stress that there exist not only $S=1$ antiferromagnetic Heisenberg chains \cite{Renard, Katsumata} but an $S=2$ chain (MnCl$_3$(bipy))\cite{Granroth} in which the presence of the Haldane gap has been experimentally confirmed. 
We give the exact form of the EE in generic VBS states in this Letter. 
Then we explicitly confirm that the part of the conjecture proposed by Vidal {\it et al}. is true for all integer-spin VBS chains. 
The relationship between the EE and the correlation function is clarified and the physical meaning of the EE in gapped models is established. We also make a comparison between the analytical results for VBS chains and the numerical results for higher-spin antiferromagnetic Heisenberg chains. The obtained results indicate that the edge-state picture is valid not only for $S=1$ Haldane chains but also for all the other integer spin-$S$ chains. 
This is a typical consequence of the non-trivial topological and/or quantum orders, where characteristic features are hidden in the bulk and appear only near the boundaries and impurities \cite{YH2}. We also discuss a potential application of the edge states as qubits for quantum computation.

Let us start with the Schwinger boson representation of generic VBS states. 
The spin operators are represented by the Schwinger bosons as 
$S_j^+ = a_j^{\dagger}b_j$, $S_j^-=b_j^{\dagger}a_j$, and 
$S_j^z=(a_j^{\dagger}a_j-b_j^{\dagger}b_j)/2$,
where $a_j\dag$ and $b_j\dag$ satisfy $[a_i,  a_j\dag]=[b_i, b_j\dag]=\delta_{ij}$ with the all the other commutators vanishing \cite{Auerbach}.
To reproduce the dimension of the spin-$S$ Hilbert space at each site, we must impose the constraint that the total boson occupation number  $a_j\dag a_j+b_j\dag b_j=2S$. 
Using the Schwinger boson representation, the spin-$S$ VBS state with two spin-$S/2$'s on the boundary is written as
\begin{equation}
|{\rm VBS}\ra=\prod_{j=0}^L (a_j^{\dagger}b_{j+1}^{\dagger}-b_j^{\dagger}a_{j+1}^{\dagger})^S|{\rm vac}\ra,
\label{VBS}
\end{equation}
where $j=1,2,...,N$ are bulk sites and $0$ and $N+1$ are end sites.
$B_{ij}\equiv a_i\dag b_j\dag-b_i\dag a_j\dag$ is a creation operator for the valence bond between $i$ and $j$ \cite{Arovas}.
The VBS state (\ref{VBS}) is a zero-energy ground state of the following Hamiltonian:
\begin{equation}
H=\sum_{j=1}^{N-1}\sum_{J=S+1}^{2S}A_J P_{j,j+1}^J+\pi_{0,1}+\pi_{N,N+1},
\nonumber
\end{equation}
where the projection operator $P_{j,j+1}^J$ projects the bond spin $\vec{J}_{j,j+1}=\vec{S}_j+\vec{S}_{j+1}$ onto the subspace of magnitude $J$. Here the coefficient $A_J$ can be an arbitrary positive value. 
The boundary terms describing interaction between spin $S/2$ and spin $S$ are explicitly written as
\begin{equation}
\pi_{0,1}=\sum_{J=S/2+1}^{3S/2}B_J P_{0,1}^J, \hspace{2mm}\pi_{N,N+1}=\sum_{J=S/2+1}^{3S/2}B_J P_{N,N+1}^J,
\nonumber
\end{equation}
with $B_J>0$.
In order to calculate reduced density matrices, it is convenient to introduce a spin coherent state. For a point $\om=(\sin\theta\cos\phi, \sin\theta\sin\phi, \cos\theta)$ on the unit sphere, the spin coherent state at each site is defined as
\begin{equation}
|\om\ra=\frac{(u a\dag+v b\dag)^{2S}}{\sqrt{(2S)!}}|0\ra,
\nonumber
\end{equation}
where $(u,v)=(\cos(\theta/2)e^{i\phi/2},\sin(\theta/2)e^{-i\phi/2})$ are spinor coodinates. Here we have already fixed the $U(1)$ gauge degree of freedom since it has no physical content.
Using $|\om\ra$, the trace of any operator $\cal O$ is written as 
${\rm Tr}{\cal O}=\frac{2S+1}{4\pi}\int d\om\la\om|{\cal O}|\om\ra$.

Let us now calculate the EE of a block of $L$ contiguous bulk spins in the VBS state (\ref{VBS}). For the density matrix of our ground state $\rho=|{\rm VBS}\ra\la {\rm VBS}|/\la {\rm VBS}|{\rm VBS}\ra$, the reduced density matrix of the block of $L$ contiguous bulk spins is defined as $\rho_L=\Tr_{\overline{\cal B}_L}\rho$. Here ${\cal B}_L$ is a block of $L$ spins and ${\overline{\cal B}}_L$ is its complement. 
The EE ${\cal S}_L=-\Tr_{{\cal B}_L}\rho_L{\rm log}_2 \rho_L$ is  determined by eigenvalues of $\rho_L$. 
Suppose that the block of $L$ contiguous spins starting from site $k$ and stretching up to $k+L-1$, where $k \ge 1$ and $k+L-1 \le N$ (Fig. 1). To obtain the reduced density matrix $\rho_L$, we take the trace over the sites $j=0, 1, ..., k-1$ and $j=k+L, ..., N, N+1$. 
Using the spin coherent state representation, $\rho_L$ is formally written as 
\begin{widetext}
\begin{equation}
\rho_L=\frac{
\int\big(\prod_{j\in {\overline{\cal B}}_L}\frac{d\om_j}{4\pi}\big)
\prod_{j=0}^{k-1} \big(\frac{1-\om_j\cdot\om_{j+1}}{2}\big)^S
\prod_{l=k+L}^{N}\big(\frac{1-\om_l\cdot\om_{l+1}}{2}\big)^S
\hspace{2mm}
Q_k\dag P_{k+L-1}\dag |{\rm VBS}_L\ra\la{\rm VBS}_L|Q_k P_{k+L-1}
}
{((2S)!)^L
\int\big(\prod_{j\in {\cal B}_L{\overline{\cal B}}_L}\frac{d\om_j}{4\pi}\big)
\prod_{j=0}^N \big(\frac{1-\om_j\cdot\om_{j+1}}{2}\big)^S
},
\label{RDM1}
\end{equation}
\end{widetext}
where boundary operators and a block of VBS state with length $L$ are defined as $Q_k=(u_{k-1}b_k-v_{k-1}a_k)^S$, $P_{k+L-1}=(a_{k+L-1}v_{k+L}-b_{k+L-1}u_{k+L})^S$ and
$|{\rm VBS}_L\ra = \prod_{j=k}^{k+L-2}(a_j\dag b_{j+1}\dag-b_j\dag a_{j+1}\dag)^S|{\rm vac}_L\ra,$
respectively.
Here we have already used the following relation: 
$\la 0| a^{S-l}b^{S+l}|\om\ra = \sqrt{(2S)!}u^{S-l}v^{S+l}$.
In Eq. (\ref{RDM1}), the integrals over $\om_{k-l-1}$ ($l=1,2,...,k-1$) can be performed by regarding $\om_{k-l}$ as a polar axis. 
The same holds for $\om_{k+L+m}$ ($m=1,2,...,N-L-k+1$). 
After integrating over these variables, we immediately notice that the reduced density matrix $\rho_L$ does not depend on both the starting site $k$ and the total length $N$. The same property for $S=1$ VBS has been proved in \cite{Korepin} 
by another approach, i.e. using the special property of maximally entangled states. The coherent state approach, however, allows us to generalize this result for more complicated cases. 
For example, we can also prove that the EE does not depend on the whole size of a VBS state on a two-dimensional Cayley tree \cite{Fan} by using the coherent state representation. 

Since the reduced density matrix does not depend on both $k$ and $N$, we can set $N=L$ without loss of generality. The following remarkable property makes it easier to calculate the EE of $L$ contiguous spins:
${\cal S}_L={\cal S}_{\hat L} \equiv -{\Tr}_{{\overline{\cal B}}_L}\rho_{\hat L}{\rm log}_2 \rho_{\hat L}$, where $\rho_{\hat L} \equiv \Tr_{{\cal B}_L}\rho$.
One can easily show this by using the Schmidt decomposition. 
Then all we have to do is to obtain the eigenvalues of the reduced density matrix of two end spin-$S/2$'s 
$\rho_{\hat L}=$
\begin{equation}
\frac{\int 
\prod_{j=1}^L\frac{d\om_j}{4\pi}
\prod_{k=1}^{L-1}\big(\frac{1-\om_k\cdot\om_{k+1}}{2}\big)^S 
P_0\dag Q_{L+1}\dag|{\rm vac}\ra\la {\rm vac}|P_0 Q_{L+1}
}
{(S!)^2
\int \big(\prod_{j=1}^L \frac{d\om_j}{4\pi}\big) \prod_{k=1}^{L-1}\big(\frac{1-\om_k\cdot\om_{k+1}}{2}\big)^S
},
\label{RDMe}
\end{equation}
where $|{\rm vac}\ra\la{\rm vac}|\equiv |0\ra_0\la 0|\otimes|0\ra_{L+1}\la 0|$. The state $P_0\dag |0\ra_0$ in the numerator of (\ref {RDMe}) is explicitly given by $(a_0\dag v_1^*-b_0\dag u_1^*)^S|0\ra$. 
From the definition of the spinor coordinates, we notice that $(u,v)$ changes to $(iv^*,-iu^*)$ when we change variables from $(\theta, \phi)$ to $(\pi-\theta,\phi+\pi)$. Then we can rewrite $P_0\dag |0\ra_0$ as
$ (-i)^S \sqrt{S!}\hspace{1mm}|-\om_1\ra_0$.
In the same way, $Q_{L+1}\dag |0\ra_{L+1}$ can be rewritten as $i^S \sqrt{S!}\hspace{1mm}|-\om_L\ra_{L+1}$. Substituting these results into Eq. (\ref{RDMe}) and changing the variables of integration from $\om_j$ to $-\om_j\hspace{1mm}(j=1,2,...,L)$, we obtain $\rho_{\hat L}=$
\begin{equation}
\frac{\int 
\prod_{j=1}^L\frac{d\om_j}{4\pi}
\prod_{k=1}^{L-1}\big(\frac{1-\om_k\cdot\om_{k+1}}{2}\big)^S 
|\om_1\ra_0\la \om_1|\otimes|\om_L\ra_{L+1} \la \om_L|
}
{\int \big(\prod_{j=1}^L \frac{d\om_j}{4\pi}\big) \prod_{k=1}^{L-1}\big(\frac{1-\om_k\cdot\om_{k+1}}{2}\big)^S}.
\label{RDMe2}
\end{equation} 
Now the physical meaning of $\rho_{\hat L}$ is quite clear. 
Eq. (\ref{RDMe2}) can be regarded as a correlation function between density matrices $|\om_1\ra_0 \la \om_1|$ and $|\om_L\ra_{L+1}\la\om_L|$. More precisely, the matrix elements of $\rho_{\hat L}$ are completely determined by the two-point correlation functions of the corresponding one-dimensional classical statistical model \cite{Arovas}. This can be checked by using the binomial expansion of $P_0$ and $Q_{L+1}$. While this interpretation enables us to understand the relation between the EE and the correlation functions, it is more convenient to use the form (\ref{RDMe2}) for the calculation of the EE.

\begin{figure}
\includegraphics[width=0.95\columnwidth,clip]{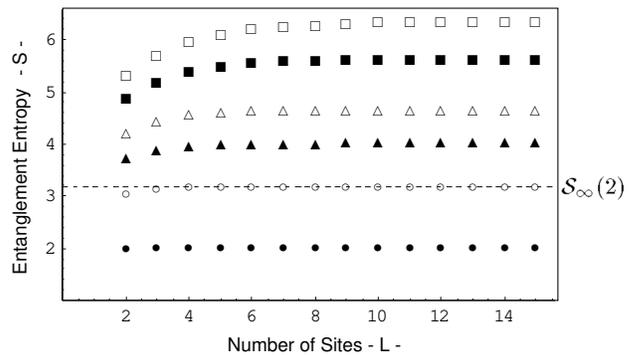}
\caption{The EE for $S=1 (\bullet)$, $S=2 (\circ)$, $S=3 (\blacktriangle)$, $S=4 (\vartriangle)$, $S=6 (\blacksquare)$, and $S=8 (\square)$ VBS chains as a function of the block size $L$. The broken line indicates the saturation value ${\cal S}_{\infty}(2)$.}
\label{graph}
\end{figure}

From now on, we follow Ref. \cite{Freitag} and obtain the eigenvalues of $\rho_{\hat L}$. In Eq. (\ref{RDMe2}), $T_{k,k+1}=(\frac{1-\om_k \cdot\om_{k+1}}{2})^S$ acts as a transfer matrix of the corresponding classical statistical model. Expanding $T_{k,k+1}$ in terms of Legendre polynomials and using the addition theorem for spherical harmonics, the transfer matrix can be rewritten as
\begin{equation}
T_{k,k+1}=\frac{4\pi}{S+1}\sum_{l=0}^S
\lambda(l)\sum_{m=-l}^l Y_l^m(\om_k)\overline{Y_l^m(\om_{k+1})}
\label{transfer}
\end{equation}
with
$\lambda(l)\equiv (-1)^l {S!(S+1)!}/[{(S-l)!(S+l+1)!}]$.
Then we substitute (\ref{transfer}) into (\ref{RDMe2}), recall the orthonormality of spherical harmonics, i.e., $\int d\om {\overline{Y_l^m(\om)}}Y_{l'}^{m'}(\om)=\delta_{ll'}\delta^{mm'}$, and obtain
\begin{equation}
\rho_{\hat L}=\frac{4\pi}{(S+1)^2}\sum_{l=0}^S \lambda(l)^{L-1} \sum_{m=-l}^l [T_l^{(m)} \otimes (T_l^{(m)})\dag],
\nonumber
\end{equation}
where irreducible $l$-th order spherical tensor operators $T_l^{(m)} \hspace{1mm}(m=-l,-l+1,...,l)$ are defined as 
$T_l^{(m)} \equiv \frac{2\cdot S/2+1}{4\pi}\int d\om |\om\ra Y_l^m(\om)\la\om|$.
We should note here that $T_l^{(m)}$ acts on the Hilbert space of the left-end spin-$S/2$ while $(T_l^{(m)})\dag$ acts on that of the right-end spin-$S/2$. Let us now introduce the following formula found in \cite{Freitag}:
\begin{equation}
\sum_{m=-l}^l [T_l^{(m)} \otimes (T_l^{(m)})\dag]=I_j(\vec{S}_0\cdot\vec{S}_{L+1}),
\label{polyn}
\end{equation}
where $\vec{S}_0$ and $\vec{S}_{L+1}$ denote the left-end and right-end spin-$S/2$'s, respectively. Here $I_j(X)$ is a $j-$th order polynomial in $X$ and determined by the following recursion relation
\begin{eqnarray}
I_{j+1}(X)=\frac{2j+3}{(S+j+2)^2}\Big(\frac{4X}{j+1}+j \Big)I_j(X) \nonumber
\\
-\frac{j}{j+1}\cdot\frac{2j+3}{2j-1}\Big( \frac{S-j+1}{S+j+2} \Big)^2 I_{j-1}(X),
\nonumber
\end{eqnarray}
with 
$I_0(X)=\frac{1}{4\pi}$, $I_1(X)=\frac{3}{4\pi}\frac{X}{(S/2+1)^2}$.
The isotropic two site tensor operators $I_j(\vec{S}_0\cdot\vec{S}_{L+1})\hspace{1mm}(j=0,1,...,S)$ are mutually orthogonal with respect to the trace inner product ${\rm Tr}_{0,L+1}(I_j I_{k})$. 
Since Eq. (\ref{polyn}) is completely determined by the polynomials in $\vec{S}_0 \cdot \vec{S}_{L+1}$, the reduced density matrix $\rho_{\hat L}$ is diagonal in the basis which diagonalizes the total spin operator  $\vec{J}_{0,L+1}=\vec{S}_0+\vec{S}_{L+1}$. Therefore, the eigenvalues of $\rho_{\hat L}$ are given by
\begin{equation}
\rho_{\hat L}(J)=\frac{4\pi}{(S+1)^2}\sum_{l=0}^S \lambda(l)^{L-1}\hspace{1mm} I_l\bigg(\frac{1}{2}J(J+1)-\frac{S}{2}\Big(\frac{S}{2}+1\Big)\bigg),
\nonumber
\end{equation}
where $J(=0,1,2, ..., S)$ is a magnitude of the total spin and each $\rho_{\hat L}(J)$ is $(2J+1)$-fold degenerate.
Finally, the EE of a block of $L$ contiguous bulk spins is explicitly written as 
\begin{equation}
{\cal S}_L=-\sum_{J=0}^S (2J+1) \rho_{\hat L}(J) \hspace{.5mm} {\rm log}_2 \hspace{.5mm} \rho_{\hat L}(J).
\label{EEeq}
\nonumber
\end{equation}
Since the reduced density matrix $\rho_{\hat L}$ approaches a $(S+1)^2$-dimensional identity matrix in the thermodynamic limit $L \to \infty$, we can see that ${\cal S}_L \le 2 {\rm log}_2 (S+1) \equiv {\cal S}_{\infty}(S)$ and approaches this upper bound exponentially fast in $L$. This saturation can be observed in Fig. \ref{graph}, where the EE ${\cal S}_L$ for various spin-$S$ VBS chains are plotted as a function of the block size $L$.  Here we confirm that the conjecture proposed by Vidal {\it et al.} is valid for all integer-spin VBS chains.

Next we make a comparison between the above results for the VBS chains and numerical results for the integer-spin Heisenberg models. 
Since $S=1$ systems have recently been extensively studied \cite{Hirano}, we study numerically the EE and the energy spectra of the $S=2$ Heisenberg model and its continuous deformations.
One of the simplest $S=2$ Hamiltonian which interpolates between 
these two models can be written as $H=$
\begin{equation}
\sum_{i=1}^N\vec{S}_i \cdot\vec{S}_{i+1}+\alpha \left\{
\frac{2}{9}
		(\vec{S}_i\cdot\vec{S}_{i+1})^2+ \frac{1}{63}
		(\vec{S}_i\cdot\vec{S}_{i+1})^3+\frac{10}{7} \right\}, 
\nonumber
\end{equation}
where $\alpha=0$ and $\alpha=1$ correspond to the Heisenberg model and the $S=2$ AKLT model, respectively.

The edge-state picture in general $S$ Haldane systems allows us to interpret the spectra as follows.
The low-lying $(S+1)$ multiplets have $(2S_{total}+1)$-fold degeneracy
in each sector when the system has open boundaries. 
These generalized Kennedy triplet states are almost degenerate, 
and are completely degenerate at the AKLT point.
This can be understood from the VBS picture. 
It would be valid for the Heisenberg model by 
some results from numerical calculations \cite{Miyashita, Qin}.
Let us now show that the ground state properties remain unchanged through the adiabatic continuation from the AKLT to the Heisenberg model.
\begin{figure}
\includegraphics[width=0.95\columnwidth,clip]{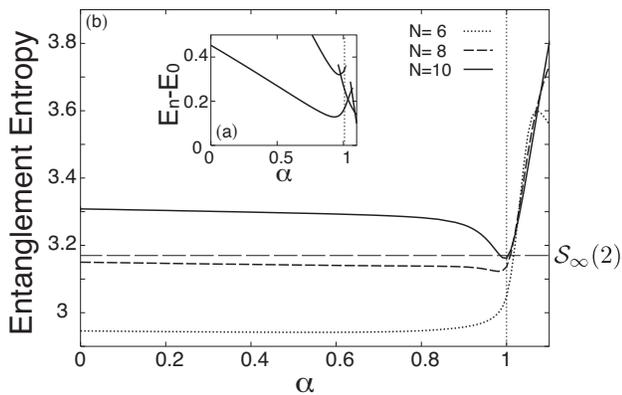}
\caption{(a)Energy gaps between the ground and the lowest two excited states in the system of $N=10$ sites with periodic boundary conditions. (b)The EE of the $S=2$ periodic $N=6, 8, 10$
Heisenberg model and its continuous deformations. }
\label{numerical}
\end{figure}
Fig. \ref{numerical} (a) shows the 
$\alpha$ dependence of the energy gaps between the ground and the lowest two excited states computed by exact diagonalizations of the system of $N=10$ sites with periodic boundary conditions. There is no level crossing
between the ground state and the first excited state, which suggests that the
low-energy behaviors of the system are adiabatically equivalent with each other in this parameter region.

Finally, let us discuss the EE in our system. The obtained results of the EE from exact diagonalizations are shown in Fig. \ref{numerical} (b).
The EE at the AKLT point has a tendency to converge to the value ${\cal S}_{\infty}(2)=2 \log_2 3 =3.16993$ as the system size increases. This value coincides with our analytically calculated one with open boundary conditions (See Fig. \ref{graph}).
The lower bound of the EE in the calculated region is given by
${\cal S}\geq2\log_2 3$, and this is the contribution from the boundaries
of the system created by taking partial trace over the subsystem.
This lower bound is equal to the EE at the AKLT
point. Taking the edge-state picture into account, we can see that this lower
bound is closely related to the number of degrees of freedom emerging at the edge.
In other words, if we can prepare a sufficiently long spin-$S$ VBS chain with open boundaries, 
each edge state behaves as a free spin-$S/2$. 
{\it This $(S+1)$-level system can be used as a qubit(qudit) for quantum computation 
by locally applying a magnetic field at the edge.} 
The EE provides a typical measure for the quantum resources.
We should note here that the EE has contributions not only from the edge state 
but also from the bulk except for the AKLT point.
In this meaning, the AKLT point is a special point since the EE has a 
contribution only from the edges created by taking partial trace. 
This fact is related to the minimum correlation length at the AKLT point.
It is also interesting that the EE at the $S=2$ AKLT point takes the minimum
value. A similar behavior has been observed in the case of $S=1$ \cite{Hirano}.
Thus, we can conjecture that the EE takes a minimum value at the AKLT point in general SU(2)-invariant models with integer-spin $S$ as far as the edge-state picture is valid.



The authors are grateful to
K. Azuma, H. Song, S. Murakami, S. Todo and N. Nagaosa for fruitful discussions.
This work was supported Grant-in-Aids (Grant No. 15104006, No. 16076205, and No. 17105002) and NAREGI Nanoscience Project from the Ministry of Education, Culture, Sports, Science, and Technology.
HK was supported by the Japan Society for the Promotion of Science. 
YH was supported by Grant-in-Aids for Scientific Research,
No. 17540347 from JSPS, No.18043007 on Priority Areas from MEXT
and the Sumitomo Foundation.


\begin{thebibliography}{99}
\bibitem{Vidal}
G. Vidal et al., Phys. Rev. Lett. {\bf 90}, 227902 (2003).
\bibitem{Levin}
M. Levin and X. G. Wen, Phys. Rev. Lett. {\bf 96}, 110405 (2006).
\bibitem{Kitaev}
A. Kitaev and J. Preskill, Phys. Rev. Lett. {\bf 96}, 110404 (2006).
\bibitem{Ryu}
S. Ryu and Y. Hatsugai, Phys. Rev. B {\bf 73}, 245115 (2006).
\bibitem{YH1}
Y. Hatsugai, J. Phys. Soc. Jpn. {\bf 74}, 1374 (2005); {\bf 75}, 123601 (2006). 
\bibitem{o3}
F. D. M. Haldane, Phys. Lett. {\bf A93}, 464 (1983).
\bibitem{Haldane}
F. D. M. Haldane, Phys. Rev. Lett. {\bf 50}, 1153 (1983).
\bibitem{AKLT1}
I. Affleck, T. Kennedy, E. Lieb, and H. Tasaki, Phys. Rev. Lett. {\bf 59}, 799 (1987).
\bibitem{AKLT2}
I. Affleck, T. Kennedy, E. Lieb, and H. Tasaki,Commun. Math. Phys. {\bf 115}, 477 (1988).
\bibitem{Verstraete and Cirac}
F. Verstraete and J. I. Cirac, Phys. Rev. A {\bf 70}, 060302(R) (2004).

\bibitem{Freitag}
W. D. Freitag and E. M${\ddot {\rm u}}$ller-Hartmann, Z. Phys. B - Condensed Matter {\bf 83}, 381 (1991).
\bibitem{Cirac}
F. Verstraete, M. A. Mart\'{\i}n-Delgado, J. I. Cirac, Phys. Rev. Lett. {\bf 92}, 087201 (2004).
\bibitem{Korepin}
H. Fan, V. Korepin, and V. Roychowdhury, Phys. Rev. Lett. {\bf 93}, 227203 (2004).
\bibitem{Renard}
J. P. Renard, M. Verdaguer, L. P. Regnault, W. A. C. Erkelens, J. Rossat-Mignod and W. G. Stirling, Europhys. Lett. {\bf 3}, 945 (1987).
\bibitem{Katsumata}
K. Katsumata, H. Hori, T. Takeuchi, M. Date, A. Yamagishi and P. Renard, Phys. Rev. Lett. {\bf 63}, 86 (1989).
\bibitem{Granroth}
G. E. Granroth {\it et al}., Phys. Rev. Lett. {\bf 77}, 1616 (1996).
\bibitem{YH2}
Y. Hatsugai, Phys. Rev. Lett. {\bf 71}, 3697 (1993).
\bibitem{Auerbach}
A. Auerbach, {\it Interacting Electrons and Quantum Magnetism}, (Springer, New York, 1998).
\bibitem{Arovas}
D, P. Arovas, A. Auerbach and F. D. M. Haldane, Phys. Rev. Lett. {\bf 60}, 531 (1988).
\bibitem{Fan}
H. Fan, V. Korepin and V. Roychowdhury, quant-ph/0511150.

\bibitem{Hagiwara}
M. Hagiwara, K. Katsumata, I. Affleck, B. I. Halperin and J. P. Renard, Phys. Rev. Lett {\bf 65}, 3181 (1990).
\bibitem{Miyashita}
S. Miyashita and S. Yamamoto, Phys. Rev. B {\bf48}, 913 (1993).
\bibitem{Qin}
S. Qin, T. K. Ng and Z. B. Su, Phys. Rev. B {\bf 52}, 12844 (1995).
\bibitem{Hirano}
T. Hirano and H. Hatsugai, unpublished.
\end{thebibliography}
\end{document}